\documentstyle[12pt,epsfig]{article}
\def\beq{\begin{equation}}
\def\eeq{\end{equation}}

\def\3p0{$^3\!P_0$}

\def\PR{Phys.\ Rev.\ }
\def\PRL{Phys.\ Rev.\ Lett.\ }
\def\ZP{Z.\ Phys.\ }
\def\NP{Nucl.\ Phys.\ }
\def\PL{Phys.\ Lett.\ }
\begin{document}
%
\begin{titlepage}
\pagestyle{empty}
\vspace*{4cm}
\begin{center}
{\large\bf DENSITY CORRELATORS IN A SELF-SIMILAR CASCADE}
\vspace{1.1cm}\\
{\sc A.~Bialas}
\vspace{0.1cm}\\
and
\vspace{0.1cm}\\
{\sc J.~Czy\.zewski}
\vspace{0.3cm}\\
{\it M. Smoluchowski Institute of Physics,}\\
{\it Jagellonian University, Cracow}\footnote{
Address: ul.\ Reymonta 4, 30-059 Krakow, Poland.
E-mail: bialas@thrisc.if.uj.edu.pl; czyzewsk@if.uj.edu.pl}
\end{center}
\vspace{1.5cm}

\begin{abstract}  
Multivariate density moments (correlators) of arbitrary order are 
obtained for the 
multiplicative self-similar cascade. This result is based on the 
calculation by Greiner, Eggers and Lipa (reference \cite{Greiner98}) 
where the correlators of the {\it logarithms} of the particle densities
have been obtained. The density correlators, more suitable for comparison
with multiparticle data, appear to have even simpler form than those obtained 
in \cite{Greiner98}.
\end{abstract} 
\vspace{0.5cm}
\noindent
{\sf TPJU 5/99}\\
{\sf June 1999}
\end{titlepage}
%
%
\section {Introduction}

It has been shown recently \cite{Greiner98,Greiner98a,Eggers98} that one can 
express the
multivariate generating function for a random multiplicative cascade
process in a simple analytical form. Using such a generating function
one can calculate multivariate density moments (correlators) \cite{Bialas88}
of any order and for an arbitrary number of bins. It was also
argued in \cite{Greiner98,Greiner98a,Eggers98} that
the cumulant density correlators obtained in this way take a particularly
simple form. They can be expressed by derivatives of the generating
function describing one vertex, i.e.\ a single-step cascade. This
beautiful result, however, required the generating function and the
correlators to be expressed in terms of logarithm of the density rather
than in terms of the  density itself. Although mathematically elegant and
adequate, e.g., for studies of turbulence, such
representation is difficult to use in analysis of
multiparticle production, as it makes the comparison with the data somewhat
complicated.
The moments expressed directly in terms of the density have been also
discussed in \cite{Greiner95} but only the recursive formula describing
them was given there.

Indeed, the experimental determinantion  of the moments of particle density 
is greatly simplified by the observation  that they can be approximated
by factorial moments of the measured  particle distribution \cite{Bialas86}. 
As this procedure is not  applicable for the moments of the logarithm of 
density, the corresponding measurements must  involve much more
sophisticated methods. This difficulty is particularly acute for low
multiplicities where Poisson fluctuations of particle number can
substantially affect the estimate of the logarithm of density.

In the present note we point out that the generating function derived in
\cite{Greiner98,Greiner98a,Eggers98} 
can also serve for calculation of the standard 
density moments and correlators and thus can be used directly for analysis of
multiparticle production. The obtained analytic expressions are rather
simple: they do not even require differentiation but simply evaluation
of the generating function at integer values of its parameters. They can
be considered as a generalization to arbitrary cascades and arbitrary
correlators of the results obtained in \cite{Bialas88} for a particular case
of the $\alpha$-model. Although our formulae represent a simple
reformulation of those derived in \cite{Greiner98,Greiner98a,Eggers98}, 
we feel that they 
deserve attention because, as explained above,  they are better suited
for analysis of multiparticle data, particularly at moderate
multiplicities.

In the next section we derive directly the formula for multivariate
density moments (correlators) in a multiplicative cascade. In Section 3
we discuss the examples of one, two and three-point correlators.
Our conclusions and comments are listed in the last section.

\section {Multi-bin correlators}
We shall use the notation of Ref.~\cite{Greiner98}.
The evolution of the self-similar cascade goes as follows. 
One starts with one bin of a given width $\Delta$, which is 
characterised by some
quantity  $\epsilon$, which can be the energy dissipation density, 
the density of particles etc., depending on the particular process 
to be decsribed by the cascade model. Its value can be chosen to be 
$\epsilon=1$ without losing generality.

This bin, the mother interval, is split into two bins\footnote{
This is the case of the binary cascade. Generalization of all the
results obtained here to a cascade characterized by three- or more-fold
splitting is straightforward},
the daughter intervals.
The contents of the two daughter bins is obtained by multiplying that of the
mother interval, $\epsilon=1$, by two numbers $q_0$ and $q_1$ 
drawn from the probability distribution $p(q_0, q_1)$ which is often
called splitting function and which we consider to be symmetric,
$p(q_0,q_1)=p(q_1,q_0)$. 
The resulting densities are 
\begin{eqnarray}
\epsilon_0 & = & q_0 \epsilon = q_0 \nonumber \\
\epsilon_1 & = & q_1 \epsilon = q_1. 
\end{eqnarray}
Now the two bins become themselves the mother intervals and are subsequently 
split into two, giving after the second step the densities
\begin{eqnarray}
\epsilon_{00} & = & q_{00} q_0 \ \ \ \ \ \ \ 
\epsilon_{01}   =   q_{01} q_0                   \nonumber \\
\epsilon_{10} & = & q_{10} q_1 \ \ \ \ \ \ \ 
\epsilon_{11}   =   q_{11} q_1,
\end{eqnarray}
where the pairs of multipliers $(q_{00},q_{01})$ and $(q_{10},q_{11})$ are
again drawn independently from the probability distribution $p$, the same 
which was used at the previous step of the cascade. 

After $J$ steps, one obtains $2^J$ bins, each one addressed by the binary 
index $k_1k_2\ldots k_J$, where every $k_j$ equals 0 or 1. The density
of every bin is the product of $J$ random multipliers $q$ which follow 
the path leading from the original mother interval to this particular bin:
\beq
\epsilon_{k_1k_2\ldots k_J} = \prod_{j=1}^J q_{k_1k_2\ldots k_j}.
\eeq

Now, the probability distribution of the particular configuration of the 
densities $\epsilon_{k_1\ldots k_J}$ is a product of the splitting functions 
$p(q_{k_1\ldots k_{j-1}0},\ q_{k_1\ldots k_{j-1}1})$ taken at each branching 
point of the cascade, convoluted with the appropriate $\delta$ functions:
\begin{eqnarray}
\lefteqn{
p(\epsilon_{0\ldots 0}, \ldots ,  \epsilon_{1\ldots 1}) =}
\nonumber \\*
& &
\int\!\!
\left[ \prod_{j=1}^J \,\prod_{k_1, \ldots,k_{j-1}=0}^1 \!\!\!\!\!\!\!\!\!
       dq_{k_1\ldots k_{j-1}0} \, dq_{k_1\ldots k_{j-1}1}\,
       p(q_{k_1\ldots k_{j-1}0},\,q_{k_1\ldots k_{j-1}1}) 
\right]\!\!
\left[ \prod_{k_1,\ldots,k_J = 0}^1\!\!\!
       \delta\!\left(
            \epsilon_{k_1 \ldots k_J} - \prod_{j=1}^J q_{k_1 \ldots k_j}
               \right)
\!\right]\!\!.
\label{probability}
\end{eqnarray}

The moments of the logarithms of the densities $\epsilon_{k_1\ldots k_J}$ 
can be calculated from the generating function $Z_T$ defined 
in \cite{Eggers98}:
\beq
Z_T(\sigma_{0\ldots 0},\ldots,\sigma_{1\ldots 1}) = 
         \left<\exp\left(
          \sum_{k_1,\ldots,k_J=0}^1 \sigma_{k_1\ldots k_J} 
              \ln\epsilon_{k_1\ldots k_J}\right)\right>,
\label{ZT}
\eeq
where the averaging goes over all configurations of the densities occuring
with the probability (\ref{probability}).
$Z_T$ provides, by differentiation, the moments $K$ of
the logarithms of the densities $\epsilon$:
\begin{eqnarray}
\lefteqn{
K(t_{0\ldots 0}, \ldots , t_{1\ldots 1}) =} \nonumber \\*
&  & \left<\prod_{k_1,\ldots,k_J = 0}^1 
            ([\ln \epsilon_{k_1 \ldots k_J})]^{t_{k_1 \ldots k_J}}\right> = 
                                              \nonumber \\
& & 
\left.
{\partial^{t_{0\ldots 0}}\over \partial (\sigma_{0\ldots 0})^{t_{0\ldots 0}}}
\ldots
{\partial^{t_{1\ldots 1}}\over \partial (\sigma_{1\ldots 1})^{t_{1\ldots 1}}}
Z_T(\sigma_{0\ldots 0},\ldots,\sigma_{1\ldots 1})
\right|_{\sigma_{k_1\ldots k_J} = 0}.
\end{eqnarray}
It was shown also in \cite{Greiner98a,Eggers98} that $Z_T$ factorizes:
\beq
Z_T(\sigma_{0\ldots 0}, \ldots , \sigma_{1\ldots 1}) =
\prod_{j=1}^J \ \prod_{k_1, \ldots,k_{j-1}=0}^1
Z_T(\sigma_{k_1\ldots k_{j-1}0},\ \sigma_{k_1\ldots k_{j-1}1}),
\label{Z-factorization}
\eeq
where
\beq
Z_T(\sigma_0,\sigma_1) =
\int dq_0 dq_1 p(q_0,q_1) \exp(\sigma_0 \ln q_0 + \sigma_1 \ln q_1)
\eeq
is the binary generating function corresponding to a cascade consisting of 
only one step and
\beq
\sigma_{k_1\ldots k_j} = \sum_{l=j+1}^J \ \sum_{k_{j+1},\ldots,k_l=0}^1
                       \sigma_{k_1\ldots k_l}
\label{descendant}
\eeq
is the sum of all $\sigma$ which are descendant with respect to 
$\sigma_{k_1\ldots k_j}$.  For example, $\sigma_0 = \sigma_{00}+\sigma_{01}$, 
where $\sigma_{00}=\sigma_{000}+\sigma_{001}$ and 
$\sigma_{01} = \sigma_{010}+ \sigma_{011}$, and so on till the last step 
of the cascade.

Let us now consider the multivariate moments of the
densities $\epsilon_{k_1\ldots k_J}$ themselves:
\beq
M(s_{0\ldots 0}, \ldots , s_{1\ldots 1}) = 
\left<\prod_{k_1,\ldots,k_J = 0}^1 
            (\epsilon_{k_1 \ldots k_J})^{s_{k_1 \ldots k_J}}\right>,
\label{M_def}
\eeq
where $s_{k_1 \ldots k_J}$ can be any non-negative integer 
numbers\footnote{
The formulae below are valid for any non-negative $s_{k_1 \ldots k_J}$.
Only for integer $s_{k_1 \ldots k_J}$, however, the correlators $M$ can be
represented by the factorial correlators and thus easily estimated from
data.}.
The average is, again, taken over all configurations at the end of the cascade:
\beq
M(s_{0\ldots 0}, \ldots , s_{1\ldots 1}) =
\int 
\left[ \prod_{k_1,\ldots,k_J = 0}^1 d\epsilon_{k_1 \ldots k_J} 
       (\epsilon_{k_1 \ldots k_J})^{s_{k_1 \ldots k_J}} \right]
p(\epsilon_{0\ldots 0}, \ldots ,  \epsilon_{1\ldots 1}),
\label{M}
\eeq
Comparing (\ref{M_def}) with (\ref{ZT}) one notices that taking
the generating function $Z_T$, (\ref{ZT}), at
fixed integer points $s_{0\ldots 0},\ldots,s_{1\ldots 1}$ gives directly 
the moments $M$:
\beq
M(s_{0\ldots 0}, \ldots , s_{1\ldots 1}) =
Z_T(s_{0\ldots 0}, \ldots , s_{1\ldots 1}).
\eeq
The factorization (\ref{Z-factorization}) can thus also be reformulated 
in terms of the moments $M$:
\beq
M(s_{0\ldots 0}, \ldots , s_{1\ldots 1}) =
\prod_{j=1}^J \ \prod_{k_1, \ldots,k_{j-1}=0}^1
M(s_{k_1\ldots k_{j-1}0},\ s_{k_1\ldots k_{j-1}1}).
\label{M_result}
\eeq
The rule (\ref{descendant}) holds here similarly for the variables 
$s_{k_1\ldots k_j}$.  The hierarchy of these variables sitting at each link 
and of the binary functions $M$ corresponding to every branching point is
shown in Fig.~1 on the example of a 3-step cascade. 
The binary moments in the r.h.s.\ of Eq~(\ref{M_result}) correspond
to the binary generating function $Z_T(\sigma_0,\sigma_1)$ taken again 
at integer points $\sigma_0 = s_0$, $\sigma_1 = s_1$ and they read:
\beq
M(s_0, s_1) = \int dq_0\, dq_1\ (q_0)^{s_0} (q_1)^{s_1} p(q_0, q_1).
\eeq
We thus conclude that all the density correlators can be expressed by 
products of the binary moments $M(s_0,s_1)$. This is surely a strong 
constraint which can be tested against the experimental data
on factorial correlators \cite{Bialas88}. The 
simplest possibility of such a test is discussed in the next section.

%
\begin{figure}[ht]
\begin{center}
{\epsfig{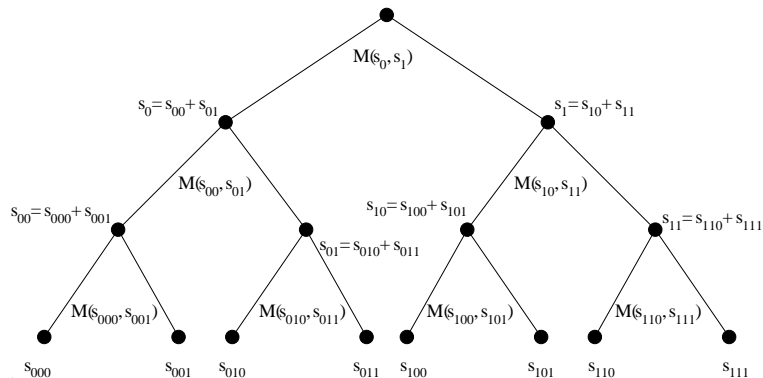}}
\label{fig1}
\parbox{13cm}{
\caption{\footnotesize
Hierarchy of the integer variables $s$ and the functions $M$ illustrating
Eq.~(\ref{M_result}).
}}
\end{center}
\end{figure}
%

\section {Examples}
Consider first the $t^{\mbox{\small\rm th}}$-order moment calculated for 
a single bin addressed by a binary number $\lambda = l_1\ldots l_J$:
\beq
M(0,0,\ldots,t,\ldots,0) = \left<(\epsilon_\lambda)^t\right>.
\eeq
This is the particular case of Eq.~(\ref{M_result}), where all the 
variables $s_\kappa$ addressed by $\kappa = k_1\ldots k_J$ different from 
$\lambda$ are set to zero, whereas $s_\lambda = t$. 
One can notice that $s_{l_1\ldots l_j} = t$ for any 
$j<J$ and $s_{k_1\ldots k_{J-j}} = 0$ for all other 
$k_1\ldots k_J \ne \lambda$.  Thus, knowing that $M(0,0) = 1$, one 
concludes that the moment $M(0,\ldots,t,\ldots,0)$ is the product of $J$ 
binary moments $M(t,0)$ or $M(0,t)$ which are all equal to each other 
due to symmetry and sit at the vertices of the trajectory leading to the 
considered bin. 
The resulting moment reads
\beq
\left<(\epsilon_\lambda)^t\right> = \left[M(t,0)\right]^J
\label{moment1}
\eeq
and it does not depend on the addres $\lambda$ of the particular bin.
Here, we recover the well-known result, namely the scaling law of the 
density moments \cite{Bialas86}.
The graphic representation of the result (\ref{moment1}) is shown in Fig.~2.

\begin{figure}[ht]
\begin{center}
{\epsfig{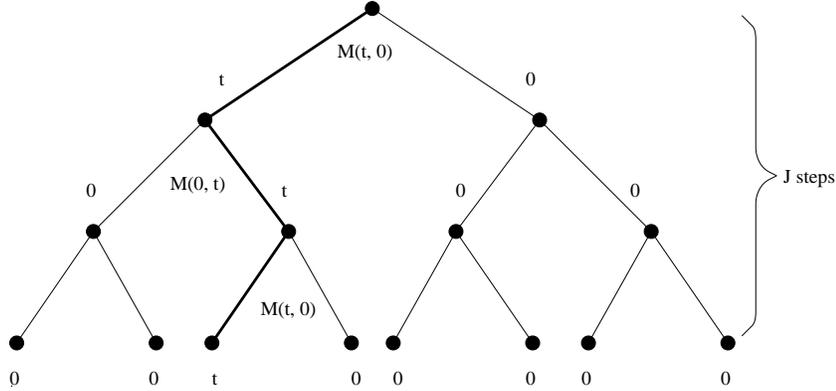}}
\label{fig2}
\parbox{13cm}{
\caption{\footnotesize
Graphic representation of the one-point moment of order $t$ for
$J=3$. The moment reads $M(0,0,t,0,0,0,0,0) =
\left< (\epsilon_{010})^t \right> = [M(t,0)]^3$.
}}
\end{center}
\end{figure}

Similarly, one can calculate the two-point correlators of order
$(t,u)$ between bins addressed by $\lambda$ and $\mu$,
\beq
M(0,\ldots,t,\ldots,u,\ldots,0) = \left< (\epsilon_\lambda)^t
                                            (\epsilon_\mu)^u \right>.
\eeq
The graphic representation of this correlator is shown in Fig.~3 and
the result is 
\beq
\left< (\epsilon_\lambda)^t (\epsilon_\mu)^u \right> = 
[M(t+u,0)]^{J-d} M(t,u) [M(t,0)M(u,0)]^{d-1},
\label{moment2}
\eeq
where $d$ is the ultrametric distance \cite{Greiner98a} between considered 
bins,
i.e.\ the number of generations leading to their common ancestor vertex.
In other words, the ultrametric distance between two bins addressed by
the binary indices $\lambda = l_1\ldots l_J$ and $\mu = m_1\ldots m_J$,
equals $d$ if
\beq
l_j = m_j\ \ \ \mbox{{\rm for}}\ \ \ \ j\le J-d
\eeq
and
\beq
l_{J-d+1} \ne m_{J-d+1}.
\eeq
%

\begin{figure}[ht]
\begin{center}
{\epsfig{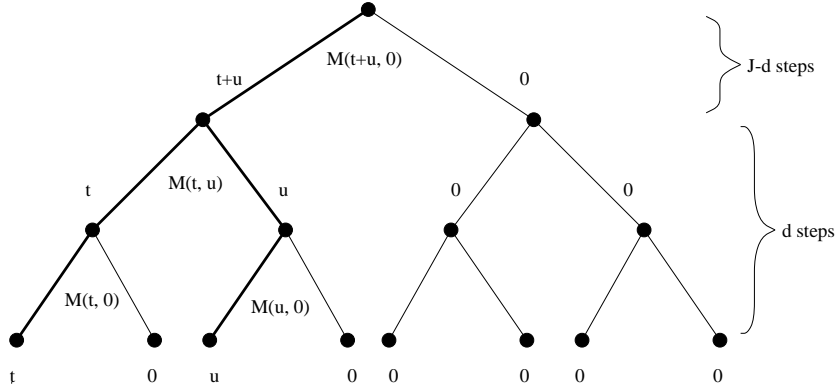}}
\label{fig3}
\parbox{13cm}{
\caption{\footnotesize
Graphic representation of the two-point correlator of order $(t,u)$ for
$J=3$ and $d=2$. The correlator is 
$M(t,0,u,0,0,0,0,0) = \left< (\epsilon_{000})^t (\epsilon_{010})^u \right> =
M(t+u,0)M(t,u)M(t,0)M(u,0)$.
}}
\end{center}
\end{figure}

Calculating the 3-point correlator is now straightforward.  Consider the
correlator of the order $(t,u,v)$ between three bins addressed by indices 
$\lambda$, $\mu$ and $\nu$, respectively.  It depends only on the 
trajectories leading to the bins. Let us assume that the ultrametric distance 
between the first two bins, $d_{\lambda\mu}$ is smaller than that of the
second pair of the bins, $d_{\mu\nu}$. This assumption does not restrict the 
generality of the solution. For such ultrametric distances between the bins,
the 3-point correlator reads:
\begin{eqnarray}
\lefteqn{
\left< (\epsilon_\lambda)^t (\epsilon_\mu)^u (\epsilon_\nu)^v \right> = 
} \nonumber \\*
& & [M(t+u+v,0)]^{J-d_{\mu\nu}} M(t+u,v) 
    [M(v,0)]^{d_{\mu\nu} - 1} 
           [M(t+u,0)]^{d_{\mu\nu} - d_{\lambda\mu} - 1} \nonumber \\
& & \times\ M(t,u) [M(t,0)M(u,0)]^{d_{\lambda\mu} - 1}.
\label{moment3}
\end{eqnarray}

Eqs~(\ref{moment1}), (\ref{moment2}) and (\ref{moment3}) provide, at least 
in principle,
a powerful test of the cascade model. Indeed, (\ref{moment1}) allows to
determine the standard intermittency parameter from a linear fit
\beq
\ln <(\epsilon_\lambda)^t > = f_t \ln \left({\Delta \over \delta}\right),
\label{intermit1}
\eeq
where
\beq
f_t = {\ln M(t,0) \over \ln 2},
\eeq
$\Delta$ is the total width of the interval in which the measurement is
taken and $\delta$ is the width of the bins into which the interval 
$\Delta$ has been divided.

Similarly, from (\ref{moment2}) one obtains the relation:
\beq
{1 \over \ln 2}
\ln \left[ {<(\epsilon_\lambda)^t (\epsilon_\mu)^u > \over
            <(\epsilon_\lambda)^t >< (\epsilon_\mu)^u >} \right] =
(J-d)(f_{t+u} - f_t - f_u) + (f_{t,u} - f_t - f_u)
\label{intermit2}
\eeq
where
\beq
f_{t,u} = {\ln M(t,u) \over \ln 2}.
\eeq

We see that the r.h.s.\ of Eq.~(\ref{intermit2}) depends only on $J-d$,
i.e.\ on the number of steps of the cascade leading to the splitting point
in which the trajectories of the bins $\lambda$ and $\mu$ separate.
It is thus independent of the bin width $\delta$. This feature is very
well confirmed by the data \cite{NA22_EMC_EMU}. The reason for this
is the fact that the above number of steps is not sensitive to the structure
of the cascade splittings hidden within the bins of width 
$\delta$ (this holds provided that the cascade splittings conserve the
particle density; in the case it does not, one can calculate the appropriate
corrections \cite{Ziaja99}).

Another important feature is that the correlator (\ref{intermit2}) does 
not depend on the physical distance $D$ between the bins for $d=J$. It means
that for the binary cascade this correlator stays constant for distances
$D$ larger than $\Delta /2$, its value being 
$f_{t,u}-f_t-f_u$ (For a cascade with tripple splitting it
would remain constant for $D > 2/3 \Delta$). It means that knowing the 
intermittency index $f_s$ for arbitrary $s$ and the above base-level value 
of the correlator allows one to determine $f_{t,u}$.

The dependence of the correlators on ultrametric distance $d$ is also worth 
studying. This
is not easy matter, however, because the relation of $d$ to the distance 
$D$ between the bins is not simple and, moreover, depends on the
dimensionality of the problem\footnote{
This was first explored in Ref.~\cite{PeschanskiSeixas90}.
}. Nevertheless, such an analysis seems not out of reach \cite{Czyzewski99}.

Finally, from Eq.~(\ref{moment3}) one obtains the relation:
\begin{eqnarray}
\lefteqn{
{1 \over \ln 2}
\ln \left[ {<(\epsilon_\lambda)^t (\epsilon_\mu)^u (\epsilon_\nu)^v> \over
    <(\epsilon_\lambda)^t >< (\epsilon_\mu)^u >< (\epsilon_\nu)^v >} \right] =
} \nonumber \\*
& &
(J-d_{\mu\nu}) (f_{t+u+v} - f_t - f_u - f_v) + 
(d_{\mu\nu}-d_{\lambda\mu}) (f_{t+u} - f_t - f_u) + \nonumber \\ & &
(f_{t+u,v} - f_{t+u} - f_v) + 
(f_{t,u} - f_t - f_u).
\end{eqnarray}
This relation can allow for checking the consistency of the above results
since it does not contain any new parameters. As it was noted before,
the coefficients $f_s$ and $f_{s,t}$ can be determined for arbitrary $s$ and
$t$ using the relations (\ref{intermit1}) and (\ref{intermit2}).

\section {Conclusions}
In conclusion, we have rewritten the results of 
Ref.~\cite{Greiner98,Greiner98a} in a form which seems better suited for 
application to data on multiple production. The resulting general 
formula for density correlators of any order can be directly compared to
multiparticle data on factorial correlators \cite{Bialas86,Bialas88}.
Since such a procedure was already applied to the $\alpha$ model with
encouraging results \cite{Kittel_Charlet94}, we feel that it may be
rewarding to analyze the data in the more general framework of arbitrary
cascade model. This should allow for a sensitive test of the relevance
of the cascade models to the processes of particle production and, if
they appear to be relevant, for better understanding of the structure
of the cascade and of its parameters.

\section* {Acknowledgements}
This investigation was supported in part by the KBN Grant No~2~P03B~086~14
and by subsydium FNP 1/99. 


\end{document}